\documentclass[12pt]{elsarticle}

\usepackage{multirow}
\usepackage{amssymb}
\usepackage{booktabs}
\usepackage{makecell}
\usepackage{amsmath}
\usepackage{url}
\usepackage{subfigure}

\makeatletter
\def\ps@pprintTitle{%
  \let\@oddhead\@empty
  \let\@evenhead\@empty
  \let\@oddfoot\@empty
  \let\@evenfoot\@empty}
\makeatother
\begin{document}

\begin{frontmatter}


% \author{Name\corref{cor1}\fnref{label2}}
% \affiliation{organization={},

\title{Social Life of Code: Modeling Evolution through Code Embedding and Opinion Dynamics} %% Article title

\author[label1]{Yulong~He}
\author[label1]{Nikita~Verbin}
\author[label2]{Sergey~Kovalchuk}

\affiliation[label1]{organization={St. Petersburg State University},
            addressline={University Embankment, 7/9},
            city={St. Petersburg},
            postcode={199034},
            % state={St. Petersburg},
            country={Russia}}

\affiliation[label2]{organization={ITMO University},
            addressline={Kronverkskiy Prospekt, 49},
            city={St. Petersburg},
            postcode={197101},
            % state={St. Petersburg},
            country={Russia}}

% \affiliation{organization={},%Department and Organization

%% Abstract
\begin{abstract}
Software repositories provide a detailed record of software evolution by capturing developer interactions through code-related activities such as pull requests and modifications. To better understand the underlying dynamics of codebase evolution, we introduce a novel approach that integrates semantic code embeddings with opinion dynamics theory, offering a quantitative framework to analyze collaborative development processes. Our approach begins by encoding code snippets into high-dimensional vector representations using state-of-the-art code embedding models, preserving both syntactic and semantic features. These embeddings are then processed using Principal Component Analysis (PCA) for dimensionality reduction, with data normalized to ensure comparability. We model temporal evolution using the Expressed-Private Opinion (EPO) model to derive trust matrices and track opinion trajectories across development cycles. These opinion trajectories reflect the underlying dynamics of consensus formation, influence propagation, and evolving alignment (or divergence) within developer communities — revealing implicit collaboration patterns and knowledge-sharing mechanisms that are otherwise difficult to observe. By bridging software engineering and computational social science, our method provides a principled way to quantify software evolution, offering new insights into developer influence, consensus formation, and project sustainability. We evaluate our approach on data from three prominent open-source GitHub repositories, demonstrating its ability to reveal interpretable behavioral trends and variations in developer interactions. The results highlight the utility of our framework in improving open-source project maintenance through data-driven analysis of collaboration dynamics.
\end{abstract}


% \begin{highlights}
% \end{highlights}

%% Keywords

\end{frontmatter}



\section{Introduction}
Traditional approaches to studying codebase evolution are focused mainlyantitative metrics such as code churn, bug frequency, and contribution patterns~\citep{4400153}. While these methods have yielded valuable insights, they often overlook the critical social dimension of software development – how developers influence each other's technical decisions and how these interactions shape the trajectory of a project. This area is usually investigated through analytical or simulation-based solutions~\citep{Namgay2024iccs,Alshomali2017iceb,Lucas2020Access}. However, developers' personality, individual behavior, motivation, and social interaction are investigated relatively shallowly.  This gap in understanding motivates our work to develop a more comprehensive approach that bridges the technical and social aspects of software evolution.

In this paper, we propose a new approach that combines code analysis with opinion dynamics to understand the relationship between code evolution and developer interactions. We assume that developers have personal views on the repository and interpret their changes to the code as reflections of these views. We use the Expressed and Private Opinion (EPO) model~\citep{EPOYE} from the opinion dynamics theory in our approach. By turning code snippets into vector representations, we can track technical changes. We then use Principal Component Analysis (PCA) to normalize these representations for easier analysis over time.

The main idea behind our work is to use the EPO model to explore how developer interactions affect code evolution. By fitting the EPO model to our processed data, we can extract meaningful parameters such as trust matrices and private opinion over time, revealing the underlying social dynamics that drive technical decisions in software projects.

\section{Related Work}

Recently, software engineering has changed with the use of machine learning to analyze source code. One of the key tools in this field are code embeddings ~\citep{chen2019embeddings}, which represent code as vectors in a continuous space. This has enabled various applications, including code search, clone detection, bug prediction, and automated program repair.

Code embeddings are vector representations of code that capture their semantic and syntactic properties.  Multiple languages are often used in modern software development environments.  One of the key advantages of code embeddings is their ability to generalize across different programming languages and codebases. Furthermore, code embeddings can capture subtle semantic relationships, such as the functional similarity between two code snippets, even if they are syntactically different.

The development of code embedding has been influenced by Word Embedding techniques in the field of natural language processing (NLP)~\citep{wordembedding}. Early approaches, such as Code2Vec~\citep{code2vec} and Code2Seq~\citep{code2seq}, utilized Abstract Syntax Trees (ASTs)~\citep{AST} to extract path information, converting code snippets into vector representations. With the rise of deep learning, pre-trained models based on the Transformer architecture~\citep{transformer}, like CodeBERT~\citep{codebert} and GraphCodeBERT~\citep{graphcodebert}, have become the most popular. These models are trained on extensive code datasets, allowing them to capture both semantic and structural nuances of code. We use the \texttt{intfloat/e5-base-v2}\footnote{\url{https://huggingface.co/intfloat/e5-base-v2}} model~\citep{e5v2}, which utilizes an optimized Transformer architecture, excels at generating efficient embeddings from large-scale datasets. 

\begin{figure}[t]
  \includegraphics[width=\columnwidth]{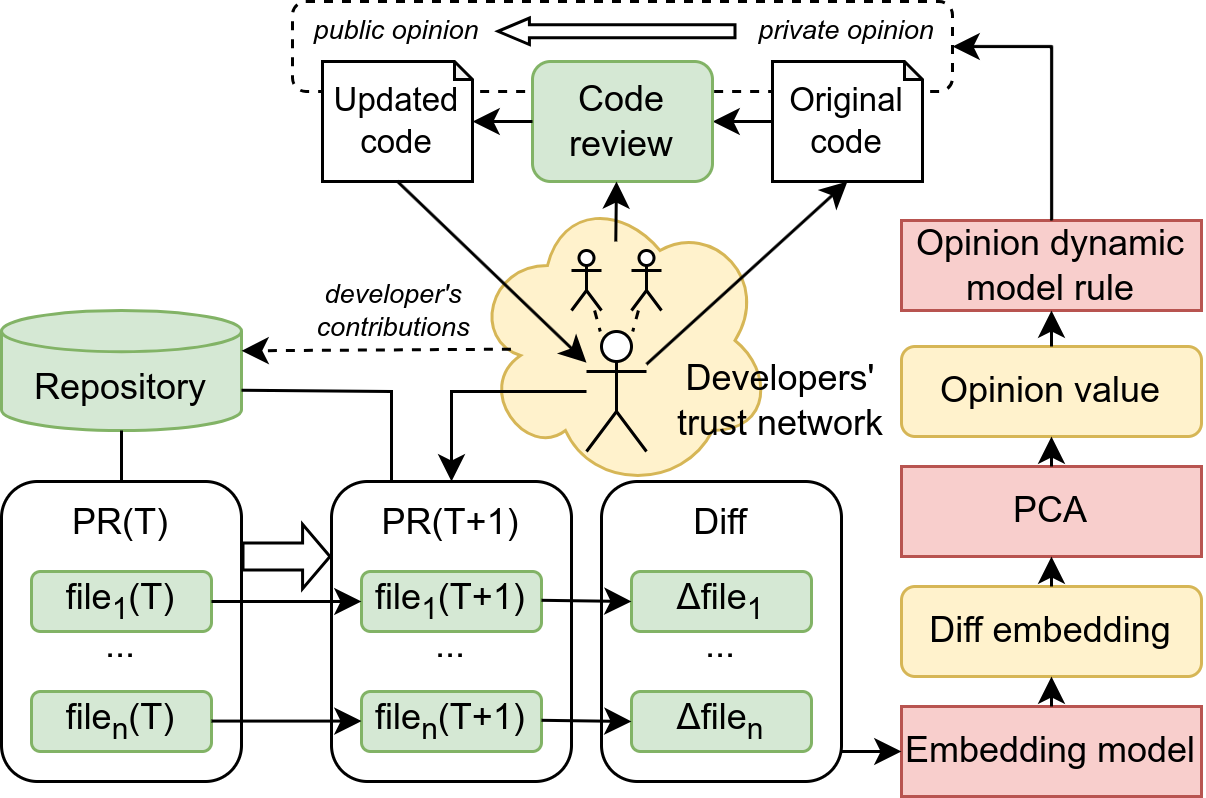}
  \caption{A general approach to opinion representation in GitHub developers' contributions}
  \label{fig:github_approach}
\end{figure}

Scholars aim to develop simple and mathematically rigorous opinion dynamic models that capture the complexity of social properties. Many classical models have been proposed, addressing the challenge of constructing effective mathematical representations for simulating the evolution of individuals' viewpoints in the real world. In 1974, DeGroot introduced the intelligence-based viewpoint evolution model (DeGroot model~\citep{DeGroot}). In this model, individuals derive their current views by assimilating the perspectives of their neighbors in the previous moment, fostering consensus within the social group. However, consensus is rarely achieved in the real world.  Friedkin et al. introduced the Friedkin-Johnsen model ~\citep{FJ}, in which an agent's subsequent opinions at each time step are influenced not only by their previous opinions but also by their initial views.  However, in practice, individuals have independent private opinions, and due to the influence of social pressure, the opinions they express are often inconsistent with their true private opinions. For this, the work of Ye et al. proposed a linear synchronous EPO model expanding from the Friedkin–Johnsen model.

Building on these foundational models, recent research has shifted further toward distinguishing expressed opinions from private beliefs, recognizing that observable convergence often masks hidden heterogeneity. In this direction, He, Proskurnikov and Sedakov’s work on sentiment evolution in Weibo blogs ~\citep{he2025opiniondynamicsmodelssentiment}
provides an important empirical and theoretical extension. Their study treats each blogger’s audience as a “macro-agent” and analyzes six months of longitudinal Weibo data, showing that emotional trajectories largely follow iterative averaging dynamics consistent with DeGroot-type models. At the same time, the authors introduce model refinements incorporating delayed perception and the distinction between private and expressed opinions, which together help explain phenomena such as emotional contagion, lagged social response, and the coexistence of superficial consensus with latent divergence. These findings highlight that even in densely connected online environments, structural delays, observational noise, and expression–private mismatches can generate complex and persistent opinion patterns that classical linear models cannot fully capture.

\section{Modeling GitHub opinion dynamics}

\subsection{GitHub opinion description}

Within our study, we consider the following approach to represent developers' social interaction (see Fig.~\ref{fig:github_approach}). Developers interact with a code repository by submitting Pull Requests (PRs), which encapsulate their contributions in the form of modified files and associated code changes (code diffs). Each PR consists of multiple files, and for each file we can extract both the original and updated code snippets. We quantify the developer's opinion on that specific file by computing the difference between the embeddings of the new and old code snippets.  We then average the opinion of each file in this PR as a snapshot of the developer's opinion at the time of the PR submission. 

When developers engage with the repository, they are exposed to the cumulative opinion expressed by all prior contributors. Developers then update their private opinion based on the trust they place in their peers' contributions. This trust-driven mechanism ensures that opinions evolve dynamically, influenced by both individual expertise and social interactions within the developer community.

The updated private opinion is subsequently expressed through new code modifications. However, the final code modification may deviate from the developers' private opinion due to other developers' interaction. For example, a junior developers' code undergoes multiple revisions based on feedback from senior developers during code review, resulting in discrepancies between the final submitted version and their initial personal drafts. Given this, we employ the EPO model that incorporates personal perspectives of the reviewers expressed in code. When a developer submits their changes, they contribute their current expressed opinion to the repository. Over time, this iterative process of reading, updating, and expressing views leads to the continuous evolution of the codebase.

\subsection{Dataset}

GitHub\footnote{\url{https://github.com/}} hosts more than 200 million repositories and connects millions of developers globally. The platform offers valuable data on PRs, code updates, and how developers interact with each other. This allows us to examine how software evolves by looking at both technical contributions and social factors. For our experimental study, we have used public dataset "88.6 Million Developer Comments from GitHub" ~\citep{88MillionDeveloperComments}. It comprises comments extracted from various GitHub collaboration channels, including commit histories, issue trackers, PR discussions, and code review processes. This dataset is organized into 135 CSV files, with three files per each of 45 popular programming languages. The files follow a standardized naming convention: \texttt{<code>\_<language>.csv}, where \texttt{<language>} represents the primary programming language; \texttt{<code>} denotes the interaction type ('\texttt{co}' for commits, '\texttt{is}' for issues, or '\texttt{pr}' for PRs).

For our analysis of code evolution patterns, we focused on PRs as they provide comprehensive documentation of code modifications, including contextual discussions and change details. Among available languages, C++ was selected due to its substantial dataset size (3,501,145 PRs), second only to Scratch - which was excluded from analysis due to its educational/non-production nature in the most cases. We then identified the most active C++ repositories based on PR count and selected the top three—swiftlang/swift, ceph/ceph, and pytorch/pytorch—for further analysis. These repositories not only accounted for the largest number of PRs (195k, 164k, and 143k, respectively) but also involved a substantial number of active contributors, ensuring that our dataset was both representative of large-scale, active development and manageable for detailed analysis. (Table~\ref{repolist}\footnote{The corresponding URL for accessing the repositories can be obtained as \texttt{https://github.com/<Name>}}). 

\begin{table}
\centering

\begin{tabular}{llll}
\hline
Name & Owner & PR    & Top   \\
     &       & count & users \\
\hline
swiftlang/swift & Apple Inc.         & 195k & 13\\
ceph/ceph       & CEPH Fdn.    & 164k & 16\\
pytorch/pytorch & PyTorch Fdn. & 143k & 28\\
  %   bitcoin        &  bitcoin&  125743\\
  %      xbmc       &      xbmc & 109686\\
  %    grpc      &       grpc  & 79832\\
  % tensorflow     &  tensorflow &  76618\\
  %     godot    &  godotengine  & 76363\\
  %     root   &  root-project  & 73166\\
  %   drake  &RobotLocomotion  & 68268\\
\hline
\end{tabular}
\caption{Summary for selected repositories}\label{repolist} % Amount of pull requests for Repositories for language 'C++'
\end{table}
We have selected top-1\% contributors according to their activity (see ``Top users'' column in Table~\ref{repolist}). This threshold ensures that our analysis focuses on highly active developers whose contributions are representative of the core development activities. To maintain consistent contributions, we required that selected developers submit PRs continuously throughout the observation period. Using these criteria, we ultimately selected seven developers from each repository who met both the activity and temporal continuity requirements. We then collected their PR IDs and retrieved the corresponding code diffs via web crawling. The final filtered dataset comprises contributions from 21 developers (seven per repository), with monthly records capturing file-level changes.

\subsection{Code embedding}

Let $\mathbf{D}$ denote the set of developers and $\mathbf{T}$ the set of time periods. For each developer $d\in \mathbf{D}$ and time period $t\in \mathbf{T}$, we define $P_d(t)$ as the set of pull requests (PRs) submitted by the developer $d$ during the time period $t$. For each PR $p$, we extract file-level code diffs, where each file modification $f\in F_p$ contains both the original and updated code snippets. Here $F_p$ is the set of files in PR $p$. We denote these paired code snippets as $(f_o,f_n)$, where $f_o$ represents the original version and $f_n$ represents the modified version of the file $f$. 

To quantify semantic changes, we employ the \texttt{intfloat/e5-base-v2}\footnote{Which is at 7-th on CoIR code embedding benchmark (https://archersama.github.io/coir/) ~\citep{li2025coircomprehensivebenchmarkcode}, This model with small model size and can get the better score on different code tasks.} code embedding model to generate vector representations for both code versions. Let $\sigma_o\in \mathbf{R}^q$ and $\sigma_n\in \mathbf{R}^q$ denote the embedding vectors for $f_o$ and $f_n$, where $q$ is the embedding dimension. 

We then compute the semantic difference vector $\sigma_f=\sigma_n-\sigma_o$. This differential embedding $\sigma_f$ captures the semantic evolution introduced by the code modifications in file $f$.

To capture the overall opinion  of a PR, we calculate the average of all vectors within that PR:
$$\sigma_p=\sum_{f\in F_p}\frac{\sigma_f}{|F_p|}$$
This averaged vector is the ``opinion of the PR'', reflecting the collective semantic direction of the changes in the PR.

Furthermore, to analyze the temporal evolution of a developer’s opinion, we aggregate all PR opinions submitted by a developer within a given time window $t$ and compute their average. This average is then defined as the developer’s opinion at time $t$:
$$\sigma_d(t)=\sum_{p\in P_d(t)}\frac{\sigma_p}{|P_d(t)|}$$
\subsection{Assessing Dimensionality Reduction Quality Metrics}
Dimensionality reduction is essential for visualizing and interpreting high-dimensional data, such as developer temporal embeddings. In this subsection, we introduce several widely used dimensionality reduction techniques and, using multiple evaluation metrics, compare their effectiveness in preserving both global and local structures.
\subsubsection{Dimensionality Reduction Techniques}

\textbf{Principal Component Analysis (PCA)} is a linear technique that identifies orthogonal directions of maximum variance in the data. By projecting data onto these principal components, PCA preserves global structure while reducing dimensionality. It is computationally efficient and particularly suitable when linear relationships dominate the data structure.

\textbf{Multidimensional Scaling (MDS)} directly optimizes the preservation of pairwise distances between data points. Unlike PCA, MDS focuses specifically on maintaining the global distance relationships from high- to low-dimensional space, making it ideal for applications where global structure interpretation is paramount, such as temporal evolution analysis.

\textbf{Independent Component Analysis (ICA)} separates a multivariate signal into additive subcomponents that are statistically independent. While similar to PCA, ICA seeks non-Gaussian components and is particularly useful when the underlying factors are assumed to be independent sources.

\textbf{Locally Linear Embedding (LLE)} preserves local linear relationships between neighboring points, effectively unfolding nonlinear manifolds while maintaining local geometry. It works particularly well when data lies on a smooth manifold with locally linear structure.

\subsubsection{Evaluation Strategies for Dimensionality Reduction}

Assessing the quality of dimensionality reduction requires multiple complementary metrics, as no single measure can capture all aspects of structure preservation.

\textbf{Global Correlation (Spearman)} measures the rank correlation between high-dimensional and low-dimensional pairwise distances. Values range from -1 to 1, with higher values indicating better preservation of global distance relationships.

\textbf{Trustworthiness} quantifies how well k-nearest neighbors in the high-dimensional space are preserved in the low-dimensional embedding~\citep{inproceedings}.

\textbf{Continuity} complements trustworthiness by measuring the consistency of neighborhood relationships in the reverse direction (from low- to high-dimensional space). It penalizes points that become neighbors in the embedding but were distant in the original space.

\textbf{Mean Relative Rank Error (MRRE)} provides a more nuanced assessment of neighborhood preservation by considering the rank differences of neighbors, offering greater sensitivity to subtle structural changes.~\citep{lee2007nonlinear}

\begin{table*}
\centering
\begin{tabular}{@{}l *{5}{c}@{}}
\toprule
\multirow{3}{*}{Repo} & \multirow{3}{*}{Method}  & \multicolumn{3}{c}{Local Structure Metrics}\\
\cmidrule(lr){3-5} 

&& \makecell{Trust\\worthiness} & Continuity & \makecell{MRRE} \\

\midrule
\multirow{4}{*}{ceph}&PCA & 0.6171 & 0.8028 & 0.20\\
&UMAP & 0.6144 & 0.8038 & 0.1987 \\
&LLE & 0.6023 & 0.7360 & 0.2649\\
&MDS & 0.5204 & 0.6569 & 0.3423 \\
\midrule
\multirow{4}{*}{pytorch}&PCA & 0.6530 & 0.7989 & 0.2064\\
&UMAP & 0.6582 & 0.7781 & 0.2252 \\
&LLE & 0.6424 & 0.7473 & 0.2552 \\
&MDS & 0.5543 & 0.6691 & 0.3292 \\
\midrule
\multirow{4}{*}{swift}&PCA & 0.7345 & 0.8752 & 0.1304  \\
&UMAP & 0.7393 & 0.8673 & 0.1378\\
&LLE & 0.6911 & 0.7921 & 0.2109\\
&MDS & 0.5752 & 0.6520 & 0.3466\\
\midrule
\multirow{4}{*}{average}&PCA&0.6682&\textbf{0.8256}&\textbf{0.1789}\\
&UMAP&\textbf{0.6706}&0.8164&0.1872\\
&LLE&0.6453&0.7585&0.2437\\
&MDS&0.5500&0.6593&0.3394\\
\bottomrule
\end{tabular}

\smallskip
\footnotesize{\textit{Note}: MRRE = Mean Relative Rank Error
% , DL = Dictionary Learning, TSVD = Truncated SVD, KPCA = KernelPCA, IPCA = IncrementalPCA, FA=FactorAnalysis
}
\caption{Performance of the Assessing Dimensionality Reduction Quality Metrics}
\label{PCA_RES}
\end{table*}

The experimental results (Table~\ref{PCA_RES}) indicate that PCA demonstrates the most stable performance in preserving local structures, particularly excelling in continuity and mean relative rank error (MRRE). This suggests that PCA is more effective in maintaining the overall structural integrity of the data in the low-dimensional space. Although UMAP shows a slight advantage in trustworthiness, its overall stability and balance are inferior to PCA. In contrast, LLE exhibits moderate performance, while MDS performs the worst across all metrics, confirming its unsuitability for this task. Among different datasets, the swift dataset achieves generally higher scores, pytorch performs relatively lower, and ceph lies in between. In summary, PCA stands out as the most suitable dimensionality reduction method, especially in scenarios requiring a balance between structural continuity and low reconstruction error.
% We can see the performance of PCA.
\begin{figure*}[htbp]
\centering  %居中
\subfigure[ceph]{   %第一张子图
\begin{minipage}{3.5cm}
\centering    %子图居中
\includegraphics[scale=0.3]{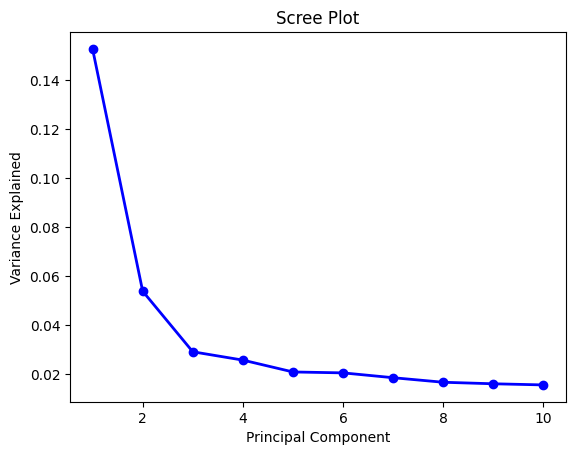}  %以pic.jpg的0.5倍大小输出
\end{minipage}
}\hspace{0.57cm}\subfigure[pytorch]{ %第二张子图
\begin{minipage}{3.5cm}
\centering    %子图居中
\includegraphics[scale=0.3]{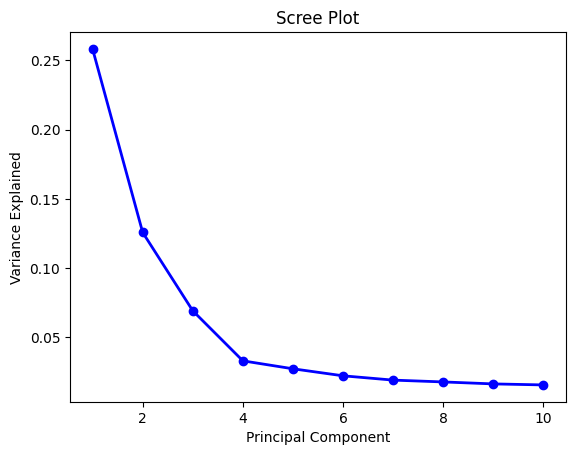}%以pic.jpg的0.5倍大小输出
\end{minipage}
}\hspace{0.8cm}\subfigure[swift]{ %第二张子图
\begin{minipage}{3.5cm}
\centering    %子图居中
\includegraphics[scale=0.3]{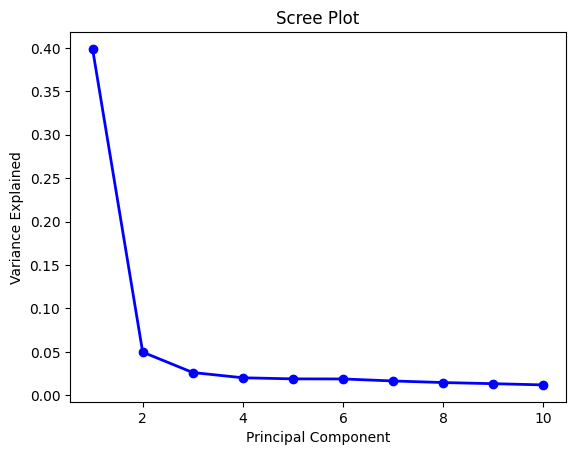}%以pic.jpg的0.5倍大小输出
\end{minipage}
}
\caption{PCA performances}    %大图名称
\label{opinions_PCA}    %图片引用标记
\end{figure*}
To assess the effectiveness of PCA, we plotted the scree plot of explained variance. The results (see Fig.\ref{opinions_PCA}) show that the first principal component accounts for the largest proportion of variance, significantly higher than subsequent components. The curve drops sharply after the first component and then flattens, indicating that the first principal component captures the dominant variation in the data. Therefore, we reduced the data to a one-dimensional representation, which preserves the essential structural information while achieving maximal dimensionality reduction.

Based on the one-dimensional representation obtained from PCA, we further apply the proposed approach to a filtered dataset consisting of seven active developers from each repository. This enables us to track the temporal evolution of opinion dynamics throughout the observed period. The variations and distances in the resulting values can then be interpreted to analyze team interactions and assess diversity within the development process.

Next, we define opinion value with Principal Component Analysis (PCA): $x_d(t)=PCA(\sigma_d(t))\in[0,1]$.
\subsection{Opinion dynamics models}

We consider the embedding of changes in the code as the "perspective" of the PR author toward the code repository. Drawing from various opinion dynamics models, we employ the EPO model, which posits that each agent possesses both a private opinion and an expressed opinion. During each iteration of opinion evolution, an agent's private opinion is initially influenced by the expressed opinions of others. Subsequently, the agent updates its private opinion and expresses its own opinion based on this updated perspective, and some other agents may give some suggestions to make them change their private opinion to final expressed opinion. We find this model fitting for code authors, as it mirrors our context. The evolution of code embedding according to the EPO model is given by:

% \begin{multline*}
% X(t+1) = diag(W)X(t) \\
% + (W-diag(W))X^e(t)
% \label{epopri}
% \end{multline*}

\begin{equation}
X(t+1) = diag(W)X(t) + (W-diag(W))X^e(t) \label{epopri}\end{equation}

\begin{equation}
X^e(t)=\Phi X(t)+(I-\Phi)A X^e(t-1)
\label{epoexp}\end{equation}

% $$
% x_d(t+1)=w_{dd}x_d(t)+\sum_{k\neq d}w_{dk}x_k^e(t),\\
% $$
% $$
% x_d^e(t)=\phi_{d}x_d(t)+(1-\phi_d)\sum_{k\neq d}a_{dk}x^e_j(t-1)
% $$
% for developer $d\in \textbf{D}$, $t\in\textbf{T}$, and initial opinion $x_d(1)$ and $x_d^e(1)$. 

% We can write in a matrix form:
Here $X(t)=(x_1(t),...,x_{|\textbf{D}|}(t))$ denote the private opinion vector and $X^e(t)=(x_1^e(t),...,x_{|\textbf{D}|}^e(t))$ the expressed opinion vector. The model incorporates two row-stochastic matrices: $W=\{w_{dk}\}$, governing how developers integrate peer assessments into their private opinions, and $A=\{a_{dj}\}$ (with zero diagonal entries, i.e., $a_{dd} = 0$), which regulates their public expression dynamics. The diagonal matrix $\Phi=diag\{\phi_1,...\phi_{|\textbf{D}|}\}$, where each $\phi_i\in [0, 1]$, modulates the interdependence between private and expressed opinions, while $I$ represents the identity matrix.

Subject to Equations \ref{epopri} and \ref{epoexp}, the decomposition $W=D+(I-D)A$ holds for some diagonal matrix $D$. We need to solve the following optimization problem:

\begin{multline*}
F(W,A,\Phi)=\sum_{t\in T}^{T-1}||x(t+1)-diag(W)x(t) \\
-(W-diag(W))x^e(t)||^2 +||x^e(t+1) \\
-\Phi x(t+1)-(I-\Phi)Ax^e(t)||^2
\end{multline*}

% $$F(W,A,\Phi)=\sum_{t\in T}^{T-1}||x(t+1)-diag(W)x(t)-(W-diag(W))x^e(t)||_2+||x^e(t+1)-\Phi x(t+1)-(I-\Phi)Ax^e(t)||_2$$

and constraints:

$$\Phi=diag\{\phi_1,...,\phi_n\},0\leq\phi_{i}\leq1$$
$$D=diag\{d_1,...,d_n\},0\leq d_{i}\leq1$$
$$diag(A)=\{0,...,0\},\sum_{j=1}^n a_{ij}=1, 0\leq a_{ij}\leq1$$
$$\sum_{j=1}^n w_{ij}=1, 0\leq w_{ij}\leq1 with\ W=D+(I-D)A$$

The problem is then solved by minimizing the objective function subject to the constraints, and the solution is validated by checking constraint satisfaction and the optimal objective value.

We have calculated code embedding 5.1 hours using GPU 'Tesla V100-PCIE-16GB', and do the optimization 3.2 hours using Mathematica on 'Lenovo ThinkPad's CPU '11th Gen Intel(R) Core(TM) i7-11370H @ 3.30GHz'.

\section{Results}
\subsection{Code Opinion Dynamic Curve}
\begin{figure*}[htbp]
\centering  %居中
\subfigure[ceph]{   %第一张子图
\begin{minipage}{3.5cm}
\centering    %子图居中
\includegraphics[scale=0.25]{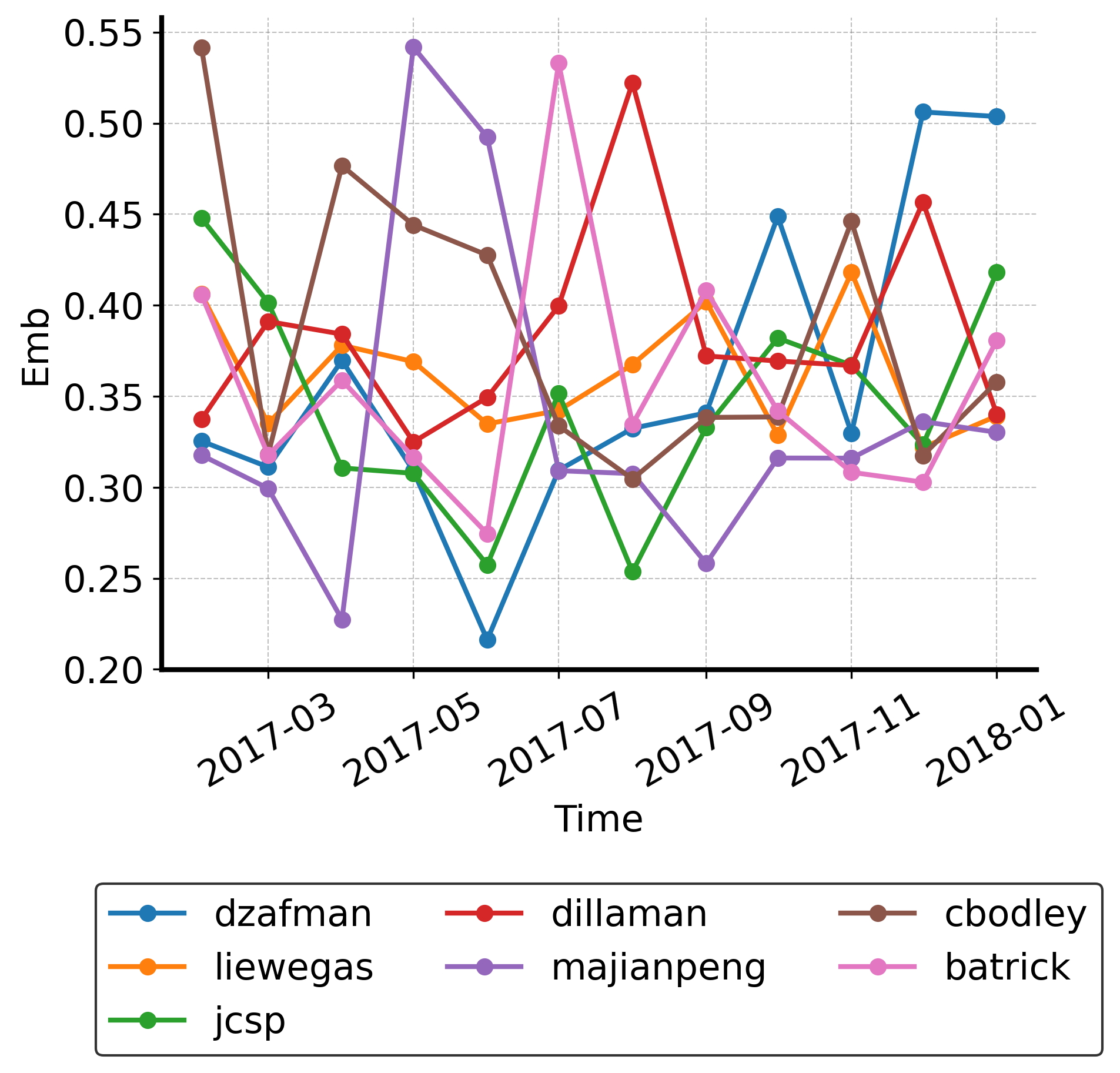}  %以pic.jpg的0.5倍大小输出
\end{minipage}
}\hspace{0.57cm}\subfigure[pytorch]{ %第二张子图
\begin{minipage}{3.5cm}
\centering    %子图居中
\includegraphics[scale=0.25]{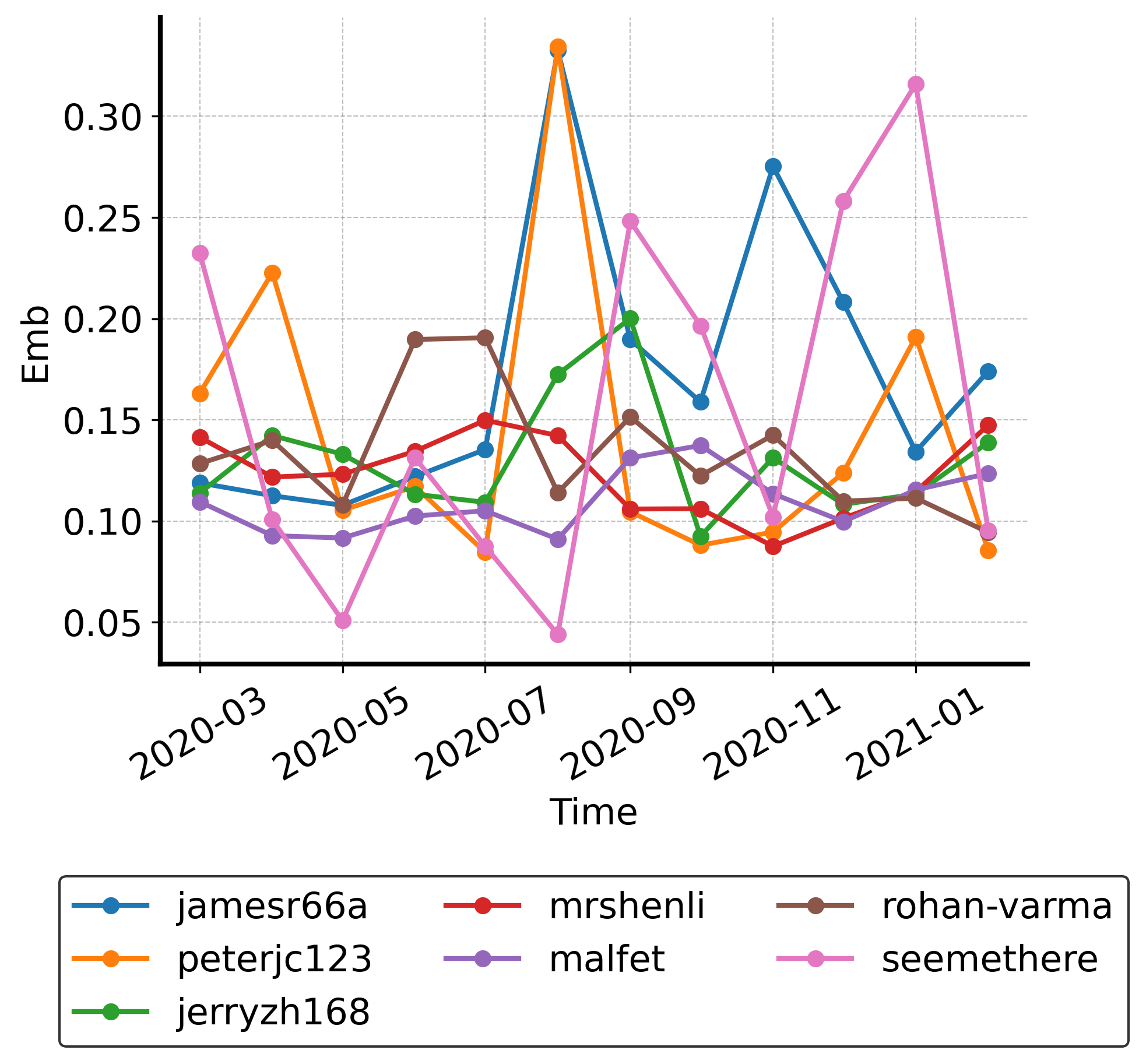}%以pic.jpg的0.5倍大小输出
\end{minipage}
}\hspace{0.8cm}\subfigure[swift]{ %第二张子图
\begin{minipage}{3.5cm}
\centering    %子图居中
\includegraphics[scale=0.25]{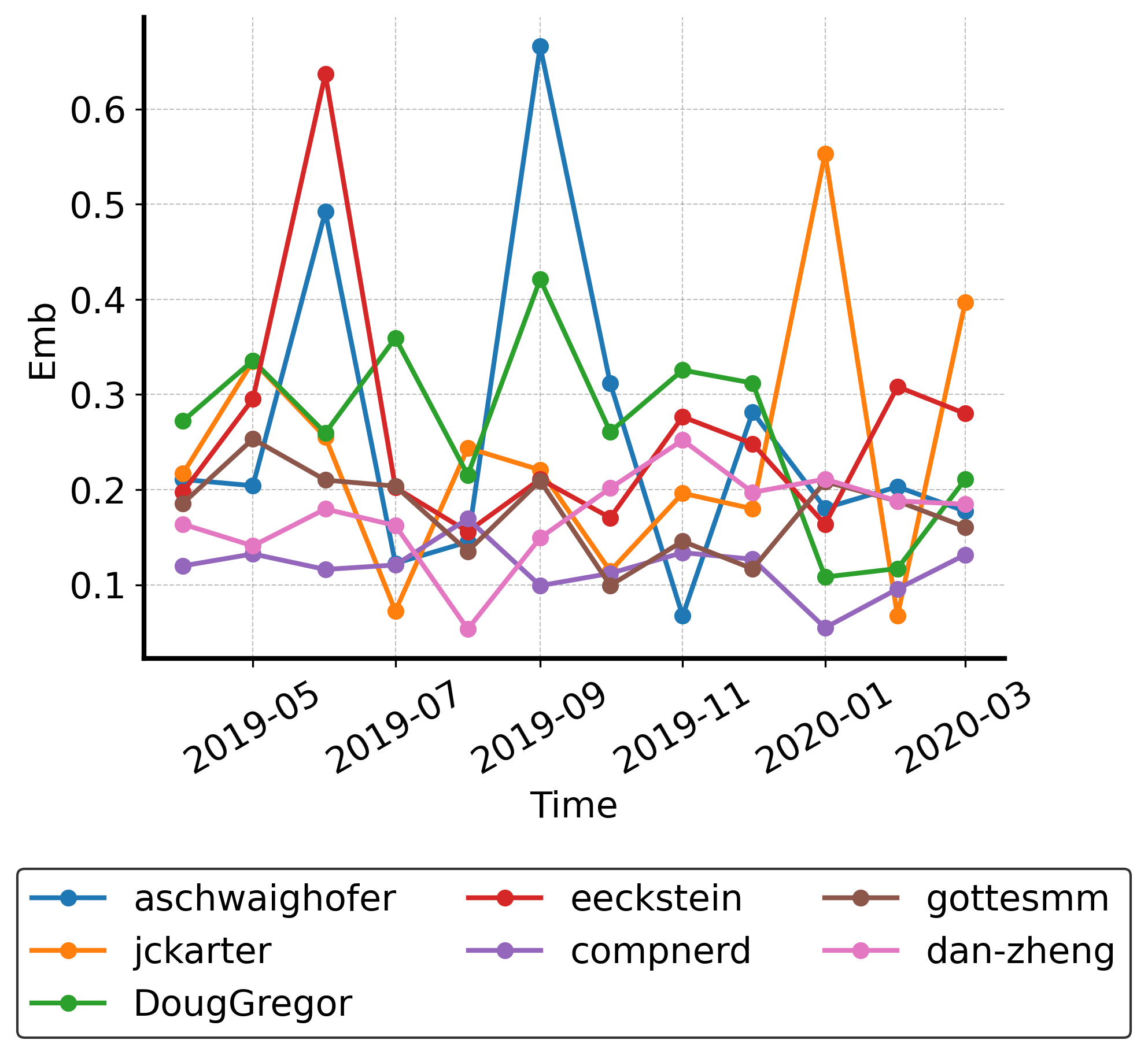}%以pic.jpg的0.5倍大小输出
\end{minipage}
}
\caption{Code “views” from the 7 most active users of repositories}    %大图名称
\label{opinions}    %图片引用标记
\end{figure*}
Our analysis of PCA-based curves representing code opinion trends in repositories (Fig.~\ref{opinions}) revealed distinct patterns: some developers' code embeddings exhibit significant fluctuations over time, while others remain relatively stable. Intuitively, minor adjustments—such as parameter tuning or comment updates—have minimal impact on the embeddings' overall meaning. In contrast, substantial modifications, like algorithmic redesigns or architectural restructuring, lead to noticeable shifts. These dynamics suggest that developers with highly variable embeddings may be senior contributors driving major changes, whereas those with stable embeddings could be junior developers focusing on incremental updates.

% During the analysis of obtained PCA-based curves representing code opinion trends in repositories (see Fig.~\ref{opinions}), we observed several interesting patterns: some developers' code embeddings swing dramatically over time, while others remain relatively stable. This makes intuitive sense—small tweaks like adjusting parameters or updating comments barely nudge the embeddings' overall meaning. But when someone overhauls core logic, such as redesigning an algorithm or restructuring the code architecture, those embeddings shift noticeably. Developers with volatile embeddings might be senior contributors driving major changes, and others who focused on smaller updates might be junior developments.

\subsection{Model Parameter Optimization and Prediction Result Analysis}

% We address the optimization problem for each of the three repositories across time periods 1–10. This process yields the opinion dynamic parameters, we use these parameters to get predict results. The results are illustrated in Fig.~\ref{prediction1-10}, where the true opinion dynamics are juxtaposed with the model's predictions. By analyzing the alignment between private and expressed opinions, we observed that the majority of developers maintain consistency in their viewpoints. They are the top 1\% of contributors with the highest code submission volumes. We think these top contributors are likely senior developers who have reached a professional level where they no longer require code review feedback from more experienced peers. There are notable trends for the developers represented by the orange curve in the repository "ceph" : Firstly, their private code opinions not only differ from publicly expressed viewpoints but also gradually converge towards the private opinions of other developers. More importantly, the increasing alignment between their private and public opinions over time demonstrates a developmental trajectory - initially showing higher receptiveness to external code modification suggestions, then progressively establishing independent judgment with experience accumulation, and ultimately reducing the adoption rate of external modification proposals. This evolutionary process visually illustrates the phased improvement of the developer's technical proficiency.

\begin{figure*}[t]
\centering
  \includegraphics[width=\textwidth]{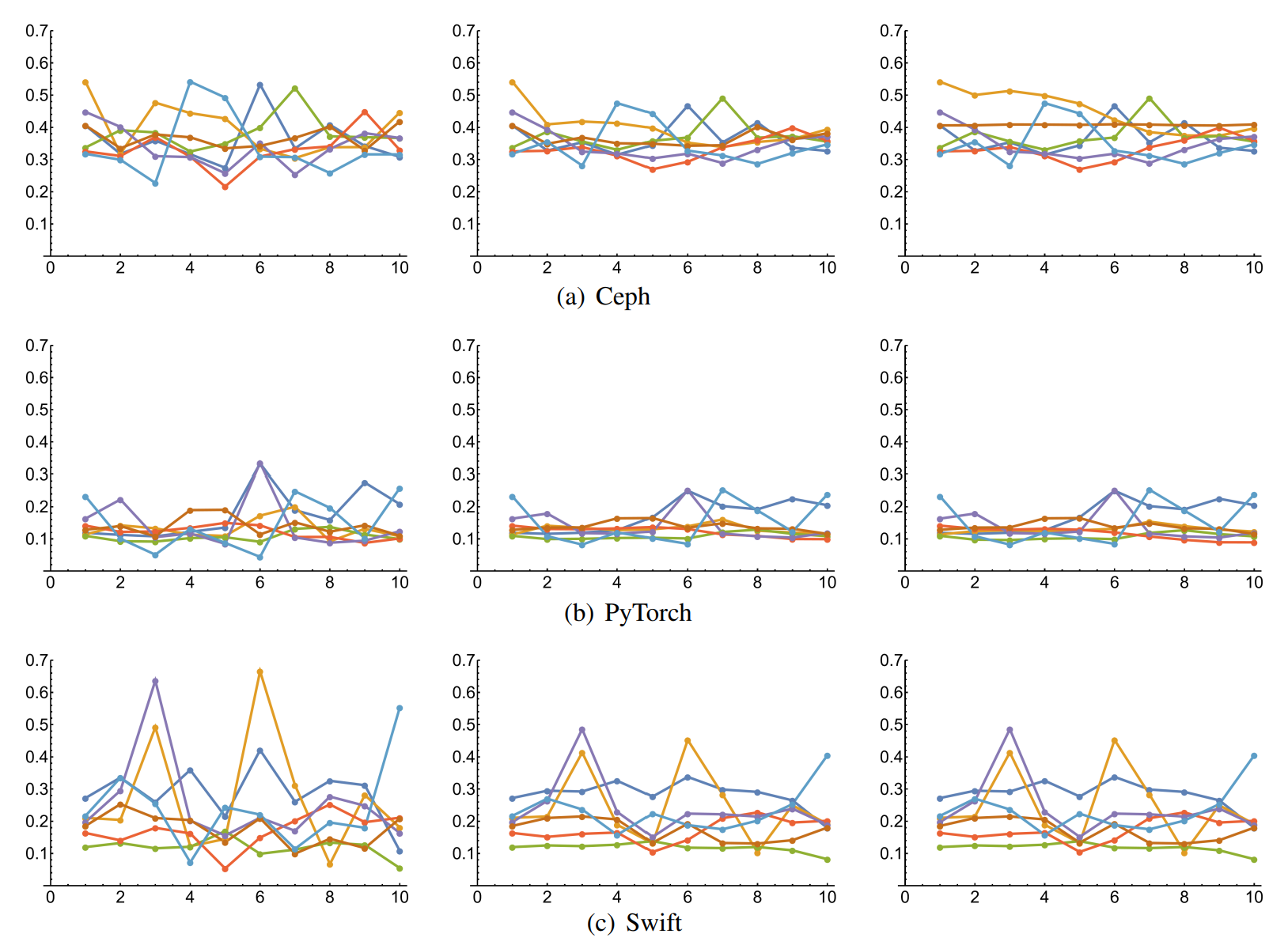}
  \caption{Comparison of true opinion datasets with predictions (in-sample). Each row corresponds to a different repository: (a) Ceph, (b) PyTorch, and (c) Swift. For each repository, the left figure represents the true opinion dataset, the center figure shows the predicted expressed opinion, and the right figure displays the predicted private opinion.}
  \label{prediction1-10}
\end{figure*}

\begin{table*}
\centering
\begin{tabular}{@{}l *{7}{c}@{}}
\toprule
Repository & \multicolumn{3}{c}{Error Metrics} & \multicolumn{3}{c}{RMSE Breakdown} \\
\cmidrule(lr){2-4} \cmidrule(lr){5-7}
& \makecell{Sum of\\Residuals} & MAE & \makecell{MAPE\\(\%)} 
& \makecell{Periods \\1--10} 
& \makecell{Periods \\11} 
& \makecell{Periods \\12}
% & \makecell{\\Periods 13}
\\
\midrule
% ceph(d)      & 0.204178 & 0.162827 & 0.170549  & 0.177106 & 0.21817 & 0.167354& 0.0810643\\
% pytorch (d)    & 0.123835 & 0.121914 & 0.302091 & 0.14376 & 0.170587 & 0.0875098 &0.13261 \\
% swift(d)       & 0.680971 & 0.264546 & 0.389682 & 0.310631 & 0.360585 & 0.236812&0.355796 \\
% \midrule
% ceph(b)      & 0.1476 & 0.0820 & 8.61  & 0.0891 & 0.2011 & 0.1614 & 0.1602\\
% pytorch (b)    & 0.1041 & 0.0654 & 16.47 & 0.0761 & 0.1602 & 0.0961 &0.1090 \\
% swift(b)       & 0.4604 & 0.1513 & 23.35 & 0.1706 & 0.3372 & 0.1053&0.2456 \\
% \midrule
ceph      & 0.1018 & 0.0733 & 7.70  & 0.0801 & 0.2264 & 0.1739 
% &0.1758 
\\
pytorch   & 0.0600 & 0.0500 & 12.70 & 0.0594 & 0.1762 & 0.0907 
% & 0.1090
\\
swift     & 0.3209 & 0.1164 & 17.73 & 0.1311 & 0.4071 & 0.1915 
% & 0.2786 
\\
\bottomrule
\end{tabular}

\smallskip
\footnotesize{\textit{Note}: MAE = Mean Absolute Error; MAPE = Mean Absolute Percentage Error; RMSE = Root Mean Square Error}
\caption{Performance of the opinion dynamics models for each repository using PCA}
\label{tab1}
\end{table*}
We address the optimization problem for each of the three repositories across ten time periods. This process yields the opinion dynamics parameters, which we then use to generate predictive results. As illustrated in Fig.~\ref{prediction1-10}, the model's predictions are juxtaposed with the true opinion dynamics. Our analysis of the alignment between private and expressed opinions reveals that most developers maintain consistency in their viewpoints. Notably, these individuals represent the top 1\% of contributors in terms of code submission volume. We hypothesize that these top contributors are likely senior developers who have attained a professional level where they no longer rely on code review feedback from more experienced peers.

In the repository 'ceph', the developers represented by the orange curve exhibit notable trends. First, their private code opinions not only diverge from their publicly expressed viewpoints but also gradually converge toward the private opinions of other developers. More importantly, the increasing alignment between their private and public opinions over time reflects a clear developmental trajectory. Initially, these developers demonstrate higher receptiveness to external code modification suggestions. With accumulating experience, they progressively develop independent judgment, ultimately reducing their adoption rate of external modification proposals. This evolutionary pattern visually illustrates the phased improvement in the developer's technical proficiency. 

\subsection{Model Validation and Dynamic Analysis of Prediction Errors}

The performance metrics for each of the three repositories are summarized in Table \ref{tab1}, offering a quantitative evaluation of the model's predictive accuracy. The model demonstrates superior performance in fitting the dynamics of the 'ceph' repository, while struggling significantly with the 'swift' repository. The opinion curves of 'swift' repository exhibits pronounced volatility, whereas the opinion trajectories of the other repositories are comparatively smoother.

\begin{figure}[t]
\centering
\includegraphics[scale=0.3]{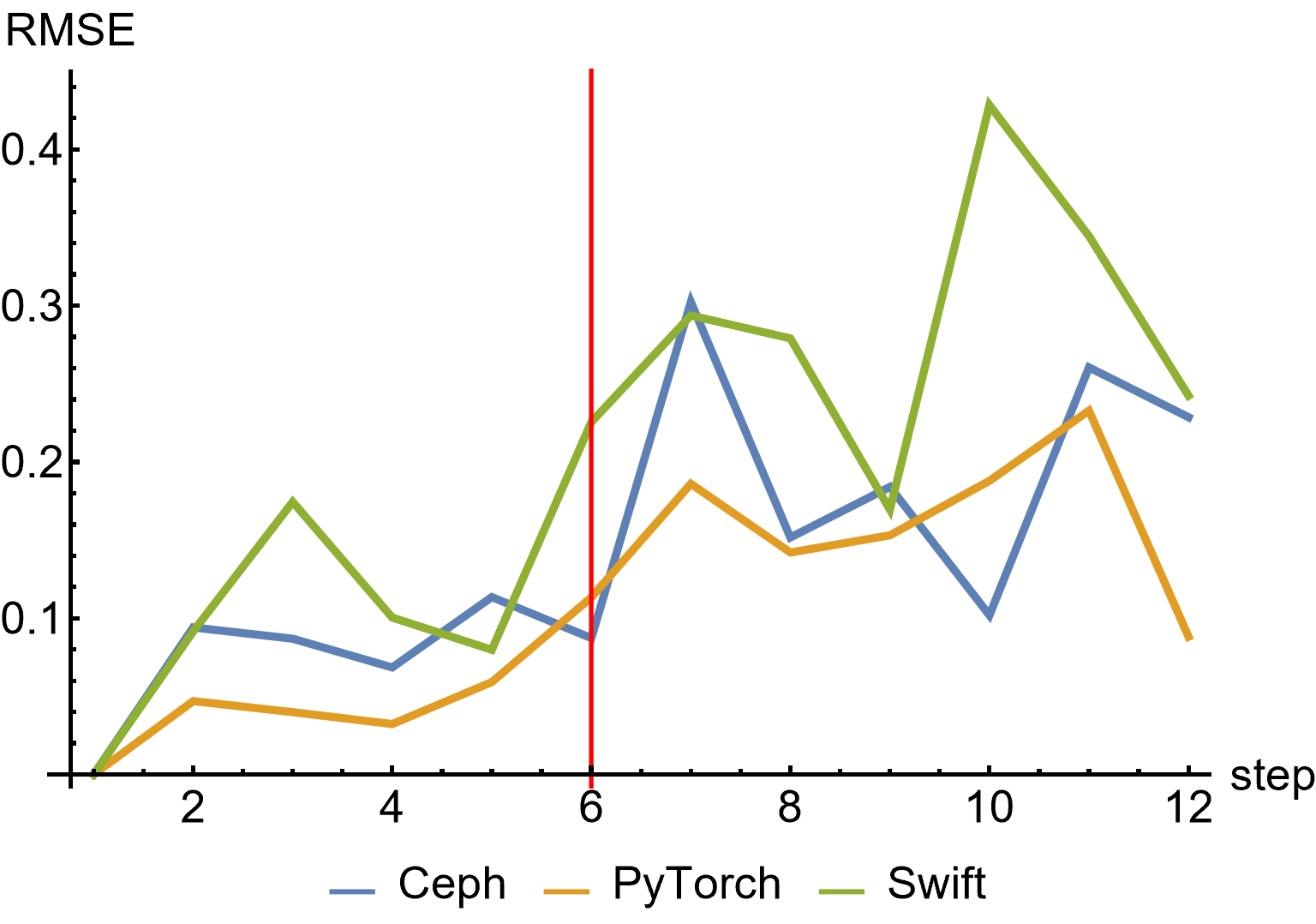}
\caption{RMSE fitting step 1-6 and predict 7-12} \label{rmse}
\end{figure}

We use parameters from optimization to predict period 11 and 12. And we can see that RMSE for period 12 consistently outperforms that of period 11 across all repositories. This observation points to an inherent hysteresis effect within the system.

\begin{figure*}[t]
\centering  %居中
\subfigure[MAE]{ %第二张子图
\begin{minipage}{5cm}
\centering    %子图居中
\includegraphics[scale=0.4]{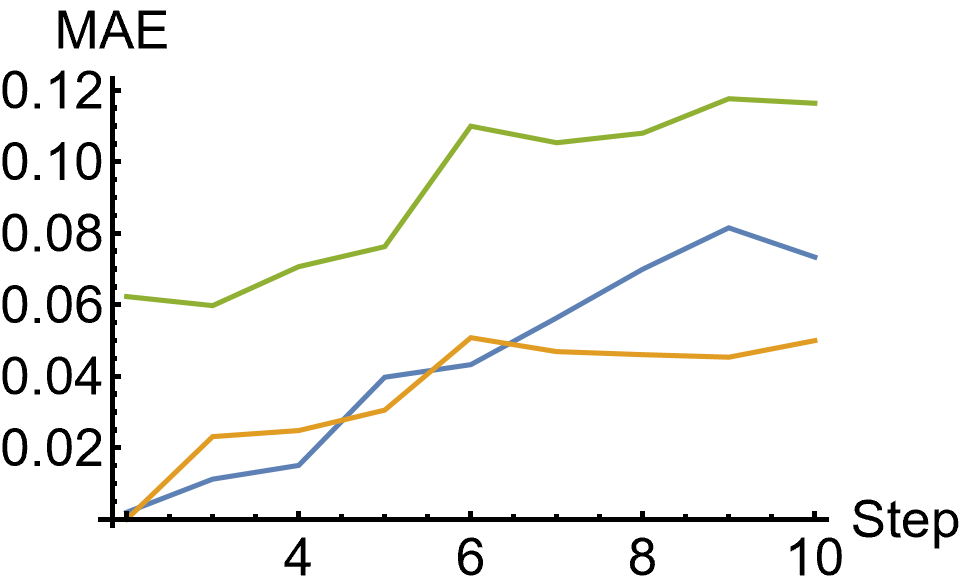}%以pic.jpg的0.5倍大小输出
\end{minipage}
}\subfigure[MAPE]{ %第二张子图
\begin{minipage}{5cm}
\centering    %子图居中
\includegraphics[scale=0.4]{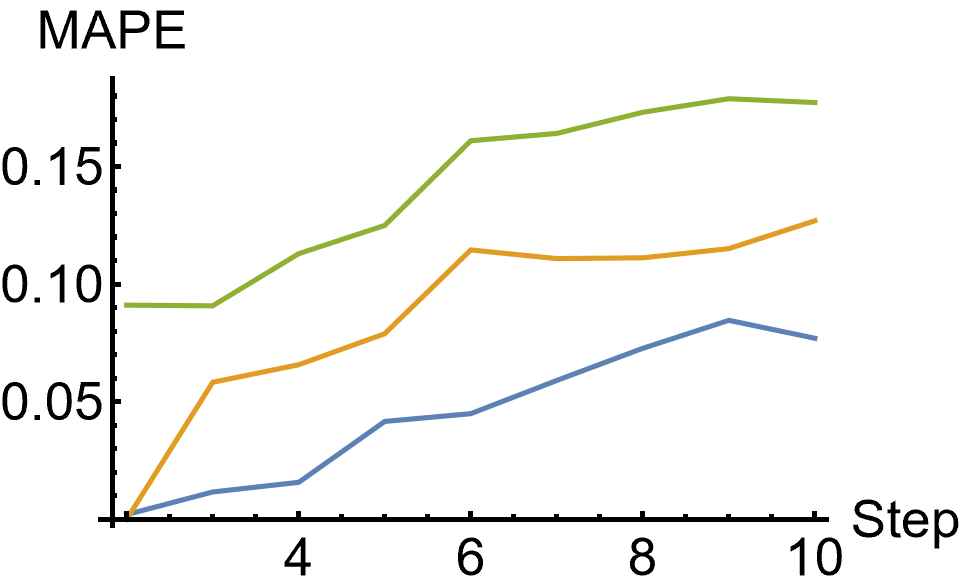}%以pic.jpg的0.5倍大小输出
\end{minipage}
}\subfigure[RMSE Period fitting]{   %第一张子图
\begin{minipage}{5cm}
\centering    %子图居中
\includegraphics[scale=0.4]{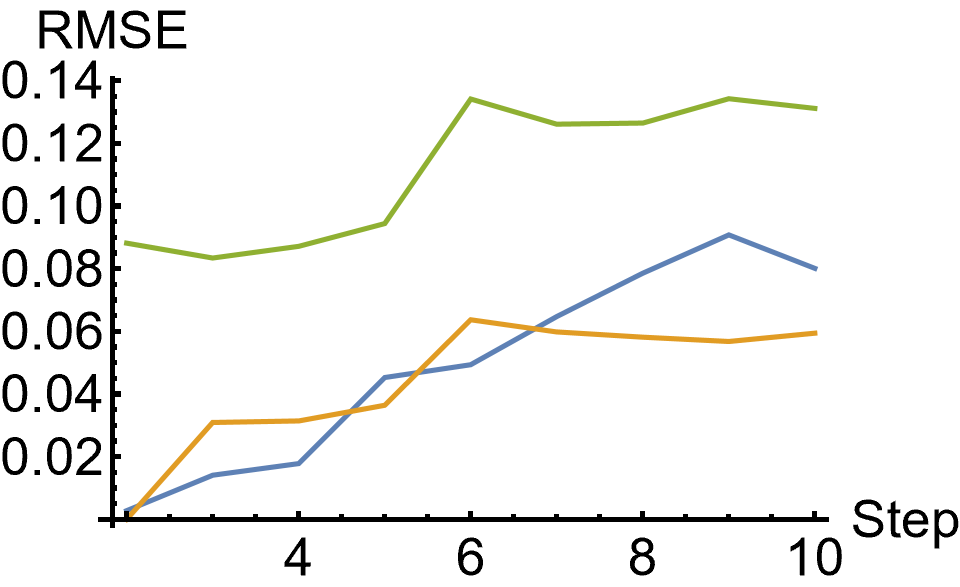}  %以pic.jpg的0.5倍大小输出
\end{minipage}
}
\subfigure[RMSE Period 11]{ %第二张子图
\begin{minipage}{5cm}
\centering    %子图居中
\includegraphics[scale=0.4]{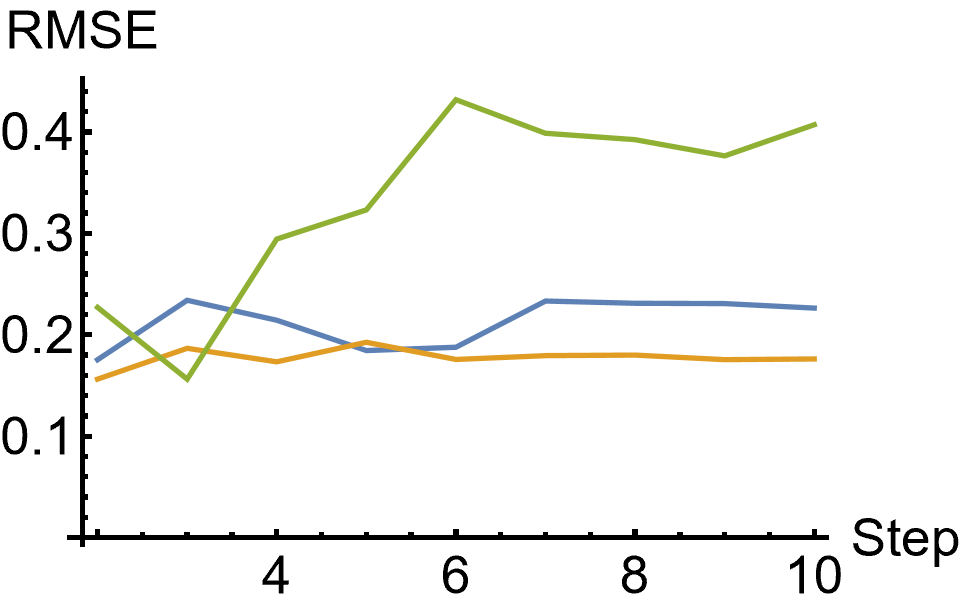}%以pic.jpg的0.5倍大小输出
\end{minipage}
}\subfigure[RMSE Period 12]{ %第二张子图
\begin{minipage}{5cm}
\centering    %子图居中
\includegraphics[scale=0.4]{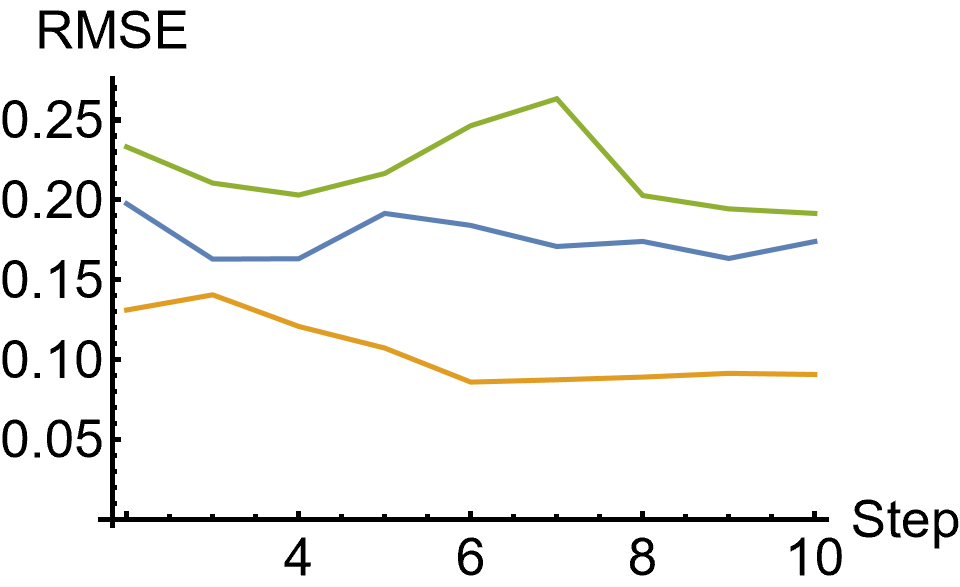}%以pic.jpg的0.5倍大小输出
\end{minipage}
}\subfigure[Sum of residuals]{   %第一张子图
\begin{minipage}{5cm}
\centering    %子图居中
\includegraphics[scale=0.44]{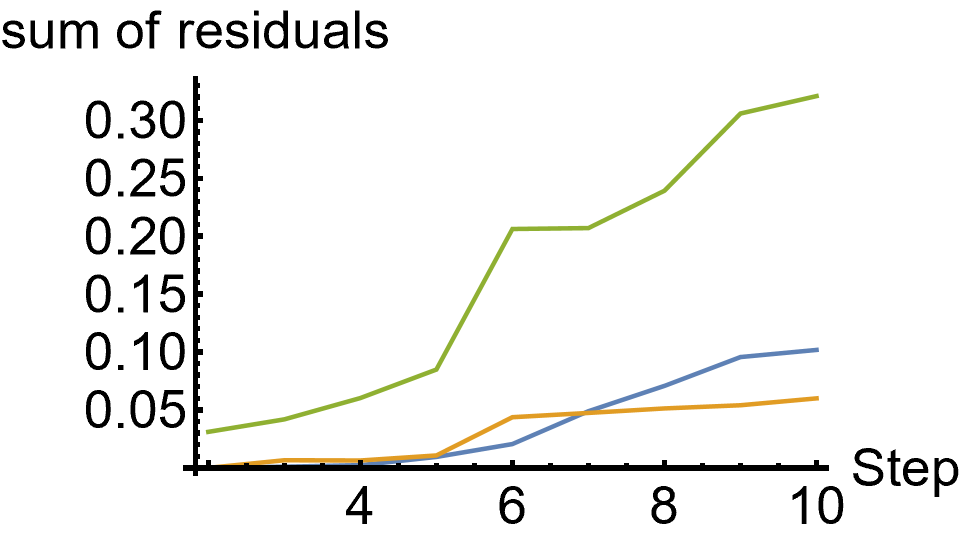}  %以pic.jpg的0.5倍大小输出
\end{minipage}
}
{   %第一张子图
\begin{minipage}{5cm}
\centering    %子图居中
\includegraphics[scale=0.44]{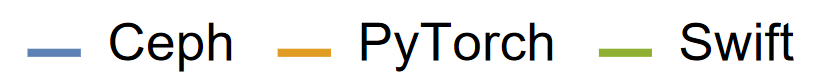}  %以pic.jpg的0.5倍大小输出
\end{minipage}
}
\caption{Performance of model fitting}    %大图名称
\label{modelfitting}    %图片引用标记
\end{figure*}

% We conduct model fitting on the final $k$ steps of the dataset, compute the correlation matrix to capture interdependencies among variables, and utilize the fitted model to generate predictions. The prediction error is then evaluated using multiple metrics—Sum of Residuals, Mean Absolute Error (MAE), Mean Absolute Percentage Error (MAPE), and RMSE -- to comprehensively assess the model's performance over periods 1–10 (see Fig.~\ref{modelfitting}). These error metrics exhibit an increasing trend as the number of steps grows.This behavior can be attributed to the growing difficult of fitting the model to larger datasets.

We perform model fitting on the final $k$ steps of the dataset, compute the correlation matrix to capture variable interdependencies, and use the fitted model to generate predictions. To evaluate prediction accuracy, we employ multiple metrics: the Sum of Residuals, Mean Absolute Error (MAE), Mean Absolute Percentage Error (MAPE), and Root Mean Squared Error (RMSE). These metrics comprehensively assess model performance across periods 1-10 (Fig.~\ref{modelfitting}). As the number of steps increases, the error metrics exhibit a rising trend, likely due to the growing difficulty of fitting the model to larger datasets.

The curve reveals that the prediction error stabilizes when a minimum six-step interval is used for model fitting. Based on this observation, we employ a six-step prediction window (time period 1–6) and compute the RMSE between the predicted and actual values for the subsequent six steps (time period 7–12; see Fig.~\ref{rmse}). The results demonstrate a gradual decline in predictive accuracy as the forecast horizon extends further into the future, a trend consistent with the inherent limitations of dynamic systems modeling.

The comparison of RMSE values across training intervals reveals that the error values in the first six training steps (time period 1-6) are consistently higher than those in the last six steps (time period 7-12). This demonstrates that developers' perspectives progressively stabilize over time as the codebase matures, reducing the necessity for frequent code modifications.

\subsection{Network analysis}

\begin{figure*}[t]
    \centering % 整体内容居中
    \subfigure[ceph]{   % 第一张子图
        \begin{minipage}{4cm}
            \centering    % 子图居中
            \includegraphics[scale=0.5]{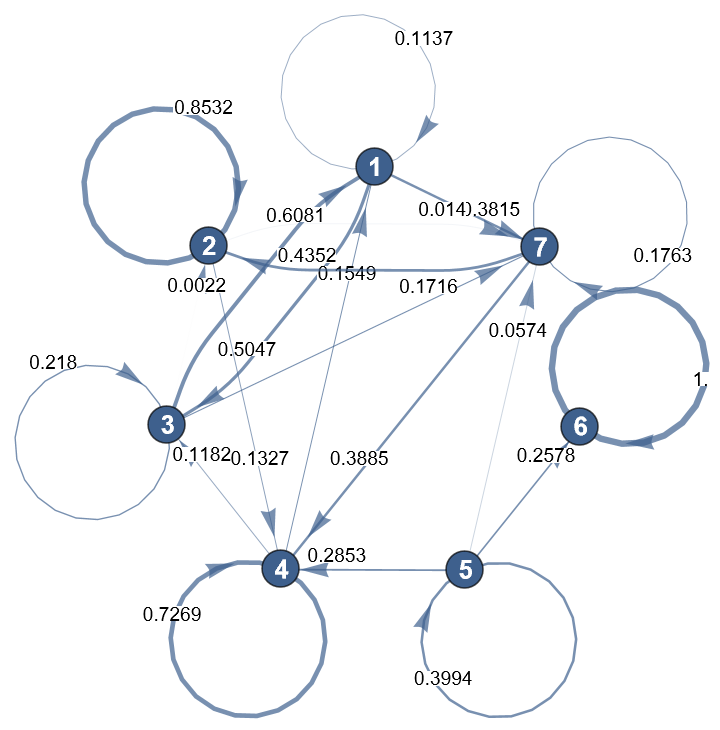}  % 图片文件
        \end{minipage}
    } \hspace{1cm} % 添加水平间距
    \subfigure[pytorch]{ % 第二张子图
        \begin{minipage}{4cm}
            \centering    % 子图居中
            \includegraphics[scale=0.5]{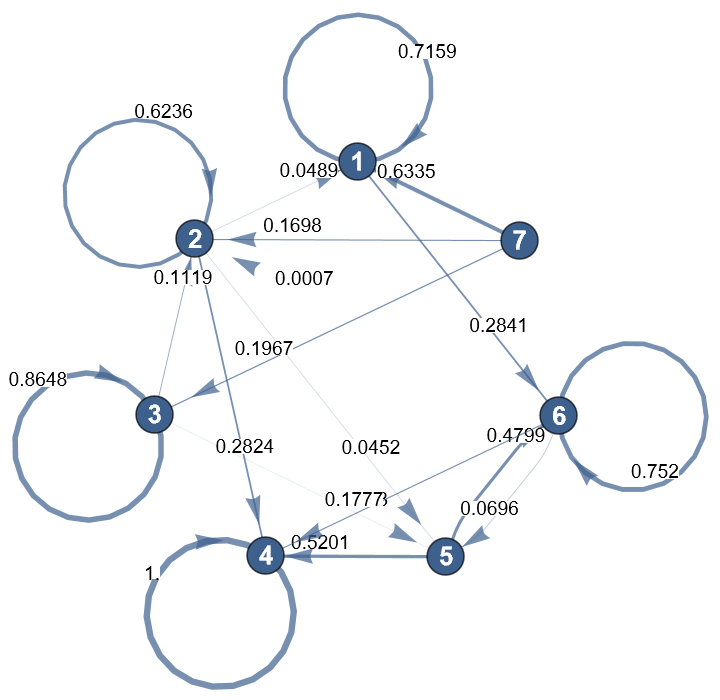} % 图片文件
        \end{minipage}
    } \hspace{1cm} % 添加水平间距
    \subfigure[swift]{ % 第三张子图
        \begin{minipage}{4cm}
            \centering    % 子图居中
            \includegraphics[scale=0.5]{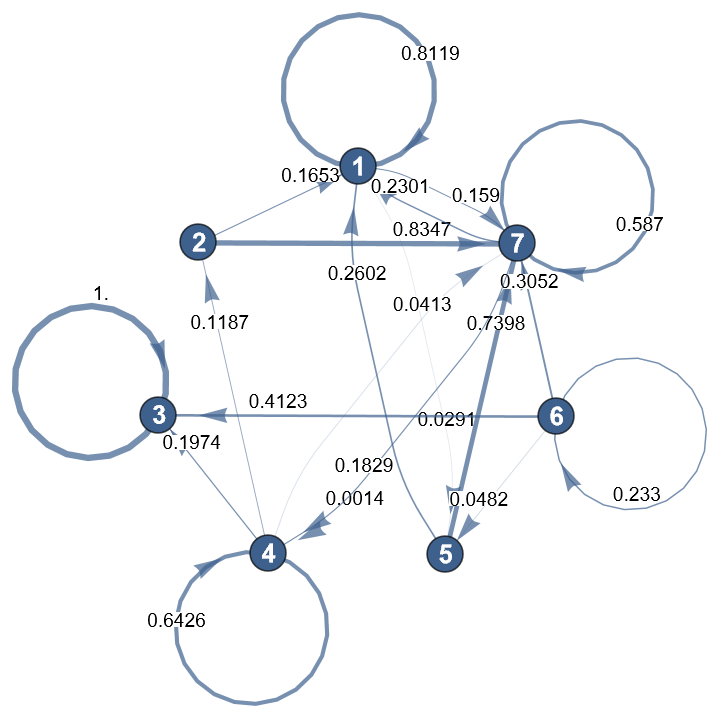} % 图片文件
        \end{minipage}
    }
    \caption{Network of 7 activate developers}    % 大图名称
    \label{networks}    % 图片引用标记
\end{figure*}

Using the matrix $W$, we can construct and analyze the dynamic system network for each repository (see Fig.~\ref{networks}), revealing insights into the influence patterns among agents. 

In the 'ceph' network, agents 2, 4, 6 show stronger independence (less influenced by others), while agent 7 is highly susceptible to peer influence. In 'pytorch', most agents are largely independent except agent 5 (fully dependent on others), with agent 7 remaining completely autonomous. The 'swift' network shows varied dynamics: agents 2, 5 fully adopt others' opinions, agent 6 maintains moderate influence balance, while others exhibit strong stubbornness (resisting external influence).

\section{Discussion and conclusion}

Our findings demonstrate that developers' views on a repository are shaped by a combination of their individual expertise and the cumulative influence of their peers. The embedding-based representation of code diffs allows us to quantify these views in a structured manner, capturing both the technical intent behind modifications and the implicit social signals embedded in the trust network. This dual perspective—technical and social—reveals the nuanced ways in which developers navigate the tension between personal beliefs and collective consensus. For instance, agents with high independence (e.g., agent 7 in the 'pytorch' network) exhibit strong autonomy, while others (e.g., agent 5 in the same network) are entirely swayed by external influences. These patterns underscore the heterogeneity in developer behavior and the role of trust in mediating opinion propagation.

The distinct patterns observed across the 'ceph', 'pytorch', and 'swift' repositories emphasize the importance of context in shaping opinion dynamics. For example, the 'ceph' network exhibits a balance between independence and influence, suggesting a relatively stable and cohesive community. In contrast, the 'swift' network displays greater volatility, with some agents fully conforming to external views while others remain rigidly independent. These differences likely stem from variations in repository size, contributor diversity, and project goals. Future research could explore how these contextual factors—such as the presence of core maintainers or the adoption of specific governance models—impact the structure and dynamics of trust networks.

In this paper, we introduced an approach to modeling opinion dynamics in GitHub repositories, offering a fresh perspective on the intricate relationship between individual contributions and collective behavior in collaborative software development. By analyzing PRs as expressions of developers' opinions, our approach uncovers how these opinions evolve over time, influenced by both technical content and social trust networks. This dual focus not only elucidates the mechanisms driving codebase evolution but also highlights the broader implications of opinion dynamics in decentralized, collaborative environments.

The presented study open a promising direction for future investigation of developers' personal behavior with opinion dynamic approach. We are planning to extend our study with more scaled investigation of the proposed solution and extension of the approach with additional factors and data sources. We suppose, integrating external factors such as issue trackers, discussion threads, and organizational policies into the analysis could provide richer insights into the interplay between code-related opinions and natural language discussions. Exploring the relationship between these elements would deepen our understanding of how developers form and express their views in collaborative software development.

\bibliographystyle{elsarticle-harv}

\bibliography{custom}

% %% The Appendices part is started with the command \appendix;
% %% appendix sections are then done as normal sections
% % \appendix
% % \section{Example Appendix Section}
% % \label{app1}

% % Appendix text.

% % %% For citations use: 
% % %%       \citet{<label>} ==> Lamport (1994)
% % %%       \citep{<label>} ==> (Lamport, 1994)
% % %%
% % Example citation, See \citet{lamport94}.

% %% If you have bib database file and want bibtex to generate the
% %% bibitems, please use
% %%
% %%  \bibliographystyle{elsarticle-harv} 
% %%  \bibliography{<your bibdatabase>}

% %% else use the following coding to input the bibitems directly in the
% %% TeX file.

% %% Refer following link for more details about bibliography and citations.
% %% https://en.wikibooks.org/wiki/LaTeX/Bibliography_Management

% % \begin{thebibliography}{00}

% % %% For authoryear reference style
% % %% \bibitem[Author(year)]{label}
% % %% Text of bibliographic item

% % \bibitem[Lamport(1994)]{lamport94}
% %   Leslie Lamport,
% %   \textit{\LaTeX: a document preparation system},
% %   Addison Wesley, Massachusetts,
% %   2nd edition,
% %   1994.

% % \end{thebibliography}
\end{document}